# Quadrantids 2008 and 2009: Detection of Dust in the Atmosphere by Polarization Twilight Sky Measurements

Oleg Ugolnikov[1], Igor Maslov[1]

[1] Space Research Institute, Russian Academy of Sciences, Profsoyuznaya st., 84/32 117997 Moscow, Russia, ougolnikov@gmail.com

## Abstract

The paper contains the results of the polarization measurements of twilight sky background during the wintertime including the epoch of Quadrantids activity in January 2008 and 2009 in Crimea (Ukraine). Analysis of the twilight sky polarization behavior had shown the barely detectable depolarization effect at the scattering altitudes above 90 km right after the Quadrantids maximum. This effect can be related with the meteoric dust in the upper atmosphere of the Earth.

## *1. Introduction*

Twilight analysis is well-known tool for the atmospheric investigations at different altitudes (Fesenkov, 1923). During the twilight period the effective altitude of single scattering is rising with the solar zenith angle. This gives the possibility to investigate the different layers of the atmosphere separately. Twilight sky background is sensitive to the appearance of the dust and aerosol in the atmosphere. In particular, twilight background can be affected by the dust scattering in the upper atmosphere after the maxima of meteor showers, as it was noticed in (Link and Robley, 1971, Link, 1975). The presence of meteoric dust was detected by the changes of twilight background brightness or its logarithmic derivative by the solar zenith angle (Mateshvili and Rietmeijer, 2002).

The basic problem of the twilight method of atmosphere investigations is the multiple light scattering. Its contribution to the twilight background intensity is about 30-40% even during the light twilight, when the Sun is close to the horizon (Ugolnikov and Maslov, 2002). Being almost constant at the solar zenith angles up to 94 degrees, this contribution is rising after that. In the dark twilight period (solar zenith angle about 99 degrees), when the effects of meteoric dust scattering can be observed, the clear atmosphere twilight background totally consists of multiple scattered emission. It is a problem that was disregarded in a number of papers where meteor dust analysis was based on the intensity measurements of the twilight sky.

This problem can be solved, if the twilight sky polarization is also measured. Polarization data helps to estimate the multiple scattering contribution (Ugolnikov, 1999) and also to detect the dust scattering in the atmosphere. This procedure was run for tropospheric aerosol (Ugolnikov and Maslov, 2005) and stratospheric aerosol (Ugolnikov and Maslov, 2009). The polarization of dust scattering is less than the background polarization of the sky even during multiple-scattering dominated period of dark twilight. So, the polarization analysis can be also used for the meteoric dust in the mesosphere causing the depolarization effect in the twilight sky. It was done in (Ugolnikov, Maslov, 2007) for the last strong maximum of Leonids in 2002. The meteoric dust was detected above 90 km during the evening twilight of November, 21, 2002, two days after the maximum with Zenithal Hourly Rate (ZHR) about 2500 (Arlt et al., 2002). The method described in (Ugolnikov, Maslov, 2007) allowed to detect the dust scattering in spite of large contribution of moonlight background to the twilight sky (last full Moon occurred at November, 20). In this paper we use the same technique to detect the less amount of dust after the recent maxima of Quadrantids shower in 2008 and 2009.

## 2. Observations

Polarization measurements of the twilight background in the near-zenith area were conducted at the Crimean Laboratory of Sternberg Astronomical Institute (Crimea, Ukraine, 44.7°N, 34.0°E, 600 m a.s.l.) by the CCD-camera with rotating polarization filter. The observations were made in the wide yellow spectral band with effective wavelength 525 nm. The main focus of the paper are the observations during the wintertime of 2008 and 2009 including the activity epoch of Quadrantids shower.

The data on the Quadrantids activity in 2008 and 2009 on the web-site of International Meteor Organization (2009) reveal the wide maximum between 02 and 11 hrs UT at January, 4, 2008 with ZHR value around 60 and the peak between 08 and 10 hrs UT with ZHR value about 80. The activity increased in 2009, the values of ZHR around 150 were observed from 09 to 13 hrs UT at January, 3, ZHR exceeded 100 for some hours before and after that. It corresponded to the daytime in the observation place, but the radiant was high above the horizon all this time, culminating close to the zenith near 06 hrs UT and shifting down to the elevation about 35 degrees at 12 hrs UT. The conditions for the dust inflow to the mesosphere above this location were very good in both years. However, the activity is many times less than for Leonids in 2002 and the depolarization effect was expected to be quite faint.

## 3. Quadrantids dust detection

Figure 1 shows the dependencies of the twilight sky polarization at the zenith on the solar zenith angle for a number of evening twilights in January 2008. The general properties of such dependency are described in (Ugolnikov and Maslov, 2005, 2007). The polarization is almost constant during the light twilight period (zenith solar angle less than 94 degrees) and during the dark twilight period (zenith solar angle around 99-100 degrees), decreasing between these two twilight stages. It is the simple reflection of ratio change between single and multiple scattering contributions.

As it was shown in (Ugolnikov et al., 2004, Ugolnikov and Maslov, 2005, 2007, 2009), in the case of clear or stable stratosphere conditions all twilight-to-twilight polarization changes are related with the multiple scattering properties being principally the same for light and dark twilight. The values of polarization during the light and dark twilight are correlated with each other and the curves in the Figure 1 are just shifted one from another as a whole.

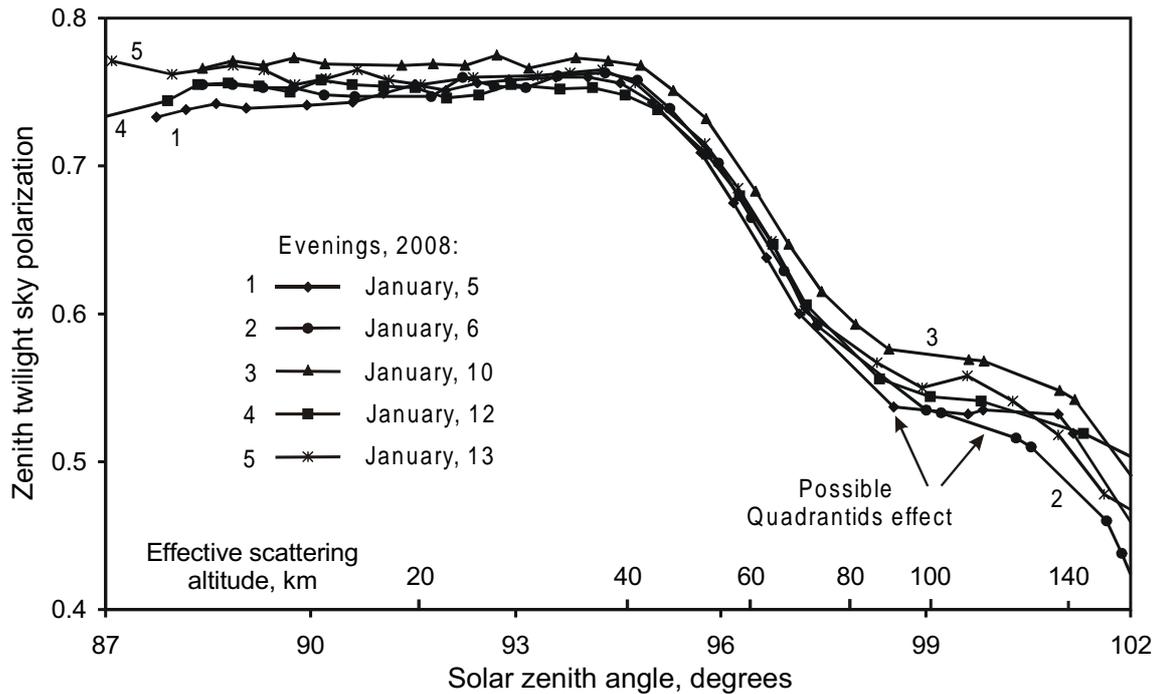

**Figure 1:** *The dependency of the zenith twilight sky polarization on the solar zenith angle end effective scattering altitude for January, 2008. Possible Quadrantids effect is shown by an arrow.*

Presence of additional dust or aerosol causes the changes of this correlation and can be detected by this way. We see that all the twilights in the Figure 1 are characterized by the close values of polarization during the light twilight period, with the little exception of the twilight of January, 10, this curve is shifted upwards. It means that the lower atmosphere conditions, where multiple scattering occurs, were the same. So, the dark twilight polarization values are expected to be the same too. But actually we see the tiny depolarization effect during the dark evening twilight of January 5 and 6, right after the Quadrantids shower maximum. This effect is shown by an arrow in the Fugure 1.

Each value of solar zenith angle corresponds to the definite effective value of single scattering altitude in the atmosphere. This value is also shown by the x-axis of the graph in the Figure 1. We see that the depolarization effect starts from the altitude about 90 km, the same was observed after the Leonids maximum in 2002. It is the layer of meteor dust moderation in the atmosphere. During the following days the dust is shifting to the lower dense atmosphere layers becoming undetectable.

During the recent Quadrantids maximum in January 2009 the ZHR was higher than for the previous one, but weather conditions were not so good, and the restricted number of observed twilights was enlighten by the Moon. The problem was solved using the scattered moonlight reduction algorithm (Ugolnikov and Maslov, 2007).

Figure 2 contains the dependency of zenith twilight sky polarization on the solar zenith angle for the evening twilight of January, 5, 2009 compared with the same ones for the clear mesosphere period in early December 2008 also with moonlight reduction and the same solar emission transfer geometry. We see that two December curves are shifted one relatively another as a whole due to the changes of multiple scattering properties, as it was described above. The curve of January, 5, 2009 is characterized by high light twilight polarization and relatively small dark twilight polarization. It can be also related with possible dark twilight depolarization caused by Quadrantids dust scattering in the mesosphere and lower thermosphere.

## 4. Conclusion

In this paper the method of polarization analysis of the twilight background was used to detect the meteoric dust in the high atmosphere after the two consecutive maxima of major meteor shower Quadrantids active during the very first days of the year. This method was used before for the storm of Leonids in 2002 (Ugolnikov and Maslov, 2007), and now the dust was more difficult to detect. However, the depolarization with the value about 0.02-0.03 was noticed for the twilights of 5-6 of January of both years. It corresponds to the contribution of meteor dust scattering to the dark twilight sky brightness not more than 0.1, if we assume the typical value of dust perpendicular scattering polarization to be about 0.2.

Such investigations can be conducted for daytime meteor showers and the ones which maximum occurs when the Sun is above the horizon at the observation place, since the meteor dust remains in the mesosphere for the several days. Moonlight scattering background can be reduced. Increase of polarization measurements accuracy (better than 0.5% for dark twilight period) will allow not only to detect but to analyze the meteoric dust scattering after the showers with maximal ZHR about 100. This will help to investigate the optical and microphysical properties of dust particles bombarding the Earth's atmosphere.

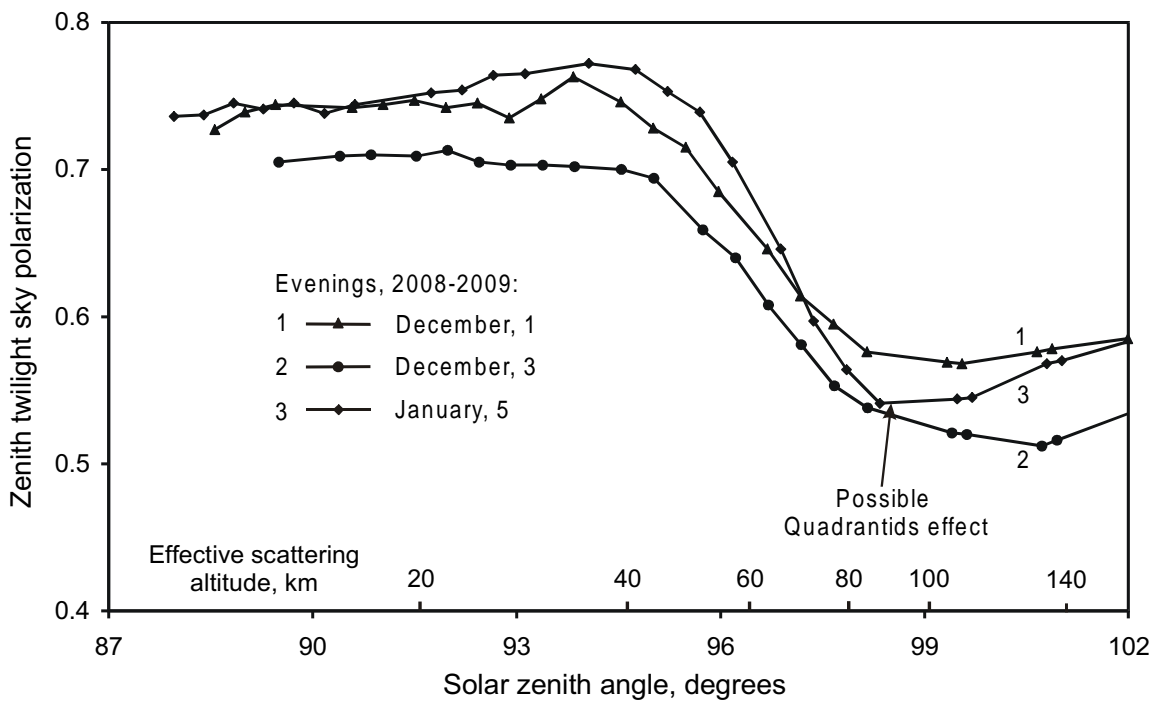

**Figure 2:** *The dependency of the zenith twilight sky polarization on the solar zenith angle end effective scattering altitude for winter 2008-2009. Moonlight scattering and night sky background are reduced. Possible Quadrantids effect is shown by an arrow.*